# A complete state-space solution model for inviscid flow around airfoils based on physics-informed neural networks

Wenbo Cao [a,b], Jiahao Song [a,b], Weiwei Zhang [a,b,*]

[a] *School of Aeronautics, Northwestern Polytechnical University, Xi'an 710072, China.*
[b] *International Joint Institute of Artificial Intelligence on Fluid Mechanics, Northwestern Polytechnical University, Xi'an, 710072, China*

**Abstract.** Engineering problems often involve solving partial differential equations (PDEs) over a range of similar problem setups with various state parameters. In traditional numerical methods, each problem is solved independently, resulting in many repetitive tasks and expensive computational costs. Data-driven modeling has alleviated these issues, enabling fast solution prediction. Nevertheless, it still requires expensive labeled data and faces limitations in modeling accuracy, generalization, and uncertainty. The recently developed methods for solving PDEs through neural network optimization, such as physics-informed neural networks (PINN), enable the simultaneous solution of a series of similar problems. However, these methods still face challenges in achieving stable training and obtaining correct results in many engineering problems. In prior research, we combined PINN with mesh transformation, using neural network to learn the solution of PDEs in the computational space instead of physical space. This approach proved successful in solving inviscid flow around airfoils. In this study, we expand the input dimensions of the model to include shape parameters and flow conditions, forming an input encompassing the complete state-space (i.e., all parameters determining the solution are included in the input). Our results show that the model has significant advantages in solving high-dimensional parametric problems, achieving continuous solutions in a broad state-space in only about 18.8 hours. This is a task that traditional numerical methods struggle to accomplish. Once established, the model can efficiently complete airfoil flow simulation and shape inverse design tasks in approximately 1 second. Furthermore, we introduce a pretraining-finetuning method, enabling the fine-tuning of the model for the task of interest and quickly achieving accuracy comparable to the finite volume method.

## 1 Introduction

In engineering problems, the solution of partial differential equations (PDEs) is often determined by multiple state parameters. It is often of interest to find the solution over a range of problem setups. Taking the example of inviscid flow around airfoils, Mach number, angle of attack and shape are all factors we care about, constituting a complete state-space, which together determine the flow. In traditional numerical methods, the flow must be solved for each specific set of state parameters independently. For instance, when assessing the performance of an airfoil, it is essential to analyze its aerodynamic characteristics under varying Mach numbers and angles of attack. To achieve an optimal airfoil for a key physical parameter, it is often necessary to simulate the flow for hundreds of different airfoil configurations. These tasks result in many repetitive tasks and expensive computational costs. The significant computational expense severely restricts the practical application of models involving PDEs in real-time predictions and many-query analysis. This limitation is particularly relevant to numerous scientific problems and real-life applications.

With the rapid development of deep learning and computing facilities, data-driven modeling methods have been used to address these challenges, enabling fast solution prediction. These methods involve various network architectures [1-5] and span a wide range of application fields [6], yielding many valuable research outcomes and demonstrating extensive application prospects. Nevertheless, data-driven modeling methods still face some limitations. The first limitation is the model training dependent on large amounts of labeled data, often computed using expensive traditional numerical methods. Since these labeled data originate from traditional numerical methods, they merely represent limited slices of data regarding the state parameters, meaning they are unevenly distributed in the joint space of spatial and state parameters. Due to this inherent flaw, the model's generalization and uncertainty must be carefully considered, constituting the second limitation. The third limitation is that once the model is established, it is difficult to fine-tune it for improving accuracy on specific tasks without the incorporation of new data.

Recently emerged methods for solving PDEs through neural network optimization, such as physics-informed neural networks (PINNs) [7], deep Ritz method [8], and deep Galerkin method [9] have been widely used to solve forward and inverse problems involving PDEs. By minimizing the loss of PDE residuals, boundary conditions and initial conditions simultaneously, the solution can be straightforwardly obtained without mesh, spatial discretization, and complicated program. With the significant progress in deep learning and computation capability, a variety of PINN-like methods have been proposed in the past few years, and have achieved remarkable results across a range of problems in computational science and engineering [10-13]. As a representative optimization-based PDE solver, PINN-like methods offer a natural method for solving PDE-constrained optimization problems, yielding a lot of valuable research outcomes, including flow visualization technology [14-17], optimal control [18-20] and inverse design for topology optimization [21]. In addition, PINN-like methods have also achieved remarkable results in solving parametric problems [9,22]. Particularly, a surrogate model has been developed for airfoil design optimization [23]. However, this surrogate model is applicable only under certain flow conditions and is limited to address flows with a Reynolds number as low as 20, thereby limiting the practical application of this expensive parametric model. Furthermore, they have no further approaches to improve the performance of the model on specific tasks.

In this paper, we focus on a typical engineering problem in aeronautics, the inviscid flow around airfoils. In our previous work [24], we presented NNfoil, a PINN method combined with mesh transformation, which successfully solved the inviscid flow around airfoils and achieved comparable accuracy to finite volume methods. In this study, we extend the input dimension of NNfoil to include all state parameters, yielding a high-dimensional parametric problem. This problem serves to highlights the advantages of PINN-like methods in solving high-dimensional parametric engineering problems. It also serves as a surrogate modeling approach, contrasting with the limitations of data-driven modeling methods previously mentioned.

The main contributions of our work can be summarized as follows:

1. We propose a complete state-space solution model designed to yield solutions for subsonic inviscid flow around airfoils across a broad spectrum of flow conditions and shapes.

2. We introduce a pretraining-finetuning method that enables the refinement of the established model on specific tasks, achieving rapid improvements in accuracy.

3. We apply this model to two engineering tasks: airfoil flow simulation and airfoil inverse design, showcasing its potential in engineering applications.

The remainder of the paper is organized as follows. In Section 2, we begin by delineating the problem setting and introducing PINNs and NNfoil. Subsequently, we present the complete state-space model and its pretraining-finetuning method. Section 3 is devoted to constructing the model and providing a detailed analysis of the results for both the pre-trained and fine-tuned models across various flows. In Section 4, the model is applied to two tasks: airfoil flow simulation and airfoil inverse design. Section 5 presents concluding remarks and directions for future research.

## 2 Methodology

2.1 Problem setting

We examine the two-dimensional Euler equation for inviscid flow, commonly employed the swift assessment of aerodynamic force. A nonconservative, dimensionless form of Euler equation is

$$\frac{\partial Q}{\partial t} = -(A\frac{\partial Q}{\partial x} + B\frac{\partial Q}{\partial y}) \tag{1}$$

where $Q = [\rho \quad u \quad v \quad p]^T$ is the vector of primitive variables; $\rho$ is the density; $u, v$ are the $x$-wise and $y$-wise components of the velocity vector $V$, respectively; $p$ is the pressure; and $A, B$ are the flux Jacobian, which have the following form

$$A = \begin{bmatrix} u & \rho & 0 & 0 \\ 0 & u & 0 & 1/\rho \\ 0 & 0 & u & 0 \\ 0 & \rho a^2 & 0 & u \end{bmatrix}, B = \begin{bmatrix} v & \rho & 0 & 0 \\ 0 & v & 0 & 0 \\ 0 & 0 & v & 1/\rho \\ 0 & 0 & \rho a^2 & v \end{bmatrix} \tag{2}$$

where $a^2 = \gamma p / \rho$ is the square of the sound speed; $\gamma = 1.4$ is the specific heat ratio. We consider the flow around airfoils. As depicted in Figure 1, the flow approaches the uniform freestream at a significant distance from the wall. Hence, the far-field boundary conditions are as follows

$$\rho_\infty = 1, u_\infty = \cos(\alpha), v_\infty = \sin(\alpha), p_\infty = 1/(\gamma Ma^2) \tag{3}$$

where $\alpha$ is the angle of attack and $Ma$ is the Mach number. As the flow cannot penetrate the wall, the velocity vector must be tangential to the surface, ensuring that the component of velocity normal to the surface is zero. Let $n$ be a unit vector normal to the surface, as depicted in Figure 1. The wall boundary condition can be written as

$$V \cdot n = 0. \tag{4}$$

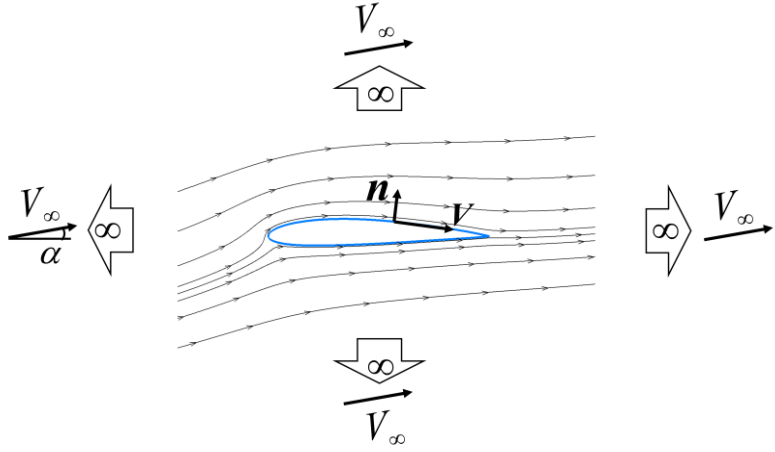

Figure 1. Boundary conditions at infinity and on the wall.

2.2 A PINN method combined with mesh transformation

In this section, we briefly introduce PINNs and NNfoil. A typical PINN employs a fully connected deep neural network (DNN) architecture to represent the solution $q$ of the dynamical system. The network takes the spatial $x \in \Omega$ and temporal $t \in [0,T]$ as the input and outputs the approximate solution $\hat{q}(x,t;\boldsymbol{\theta})$. The spatial domain typically has 1-, 2- or 3-dimensions in most physical problems, and the temporal domain may nonexistent for time-independent (steady) problems. The result of PINNs is determined by the network parameters $\boldsymbol{\theta}$, which are optimized with respect to PINNs loss function during the training process. To formulate the loss function for PINNs, we consider $q$ to be mathematically described by partial differential equations in the general form:

$$\begin{aligned}&\mathcal{N}[q(x,t)] = 0, x \in \Omega, t \in (0,T]\\&\mathcal{I}[q(x,0)] = 0, x \in \Omega\\&\mathcal{B}[q(x,t)] = 0, x \in \partial\Omega, t \in (0,T]\end{aligned} \qquad (5)$$

where $\mathcal{N}[\cdot]$, $\mathcal{I}[\cdot]$ and $\mathcal{B}[\cdot]$ are the PDE operator, the initial condition operator, and the boundary condition operator, respectively. Then PINNs loss function is defined as

$$\begin{aligned}&\mathcal{L} = \lambda_{PDE}\mathcal{L}_{PDE} + \lambda_{BC}\mathcal{L}_{BC} + \lambda_{IC}\mathcal{L}_{IC}\\&\mathcal{L}_{PDE} = \left\|\mathcal{N}[\hat{q}(\cdot;\boldsymbol{\theta})]\right\|^2_{\Omega\times(0,T]}\\&\mathcal{L}_{IC} = \left\|\mathcal{I}[\hat{q}(\cdot,0;\boldsymbol{\theta})]\right\|^2_{\Omega}\\&\mathcal{L}_{BC} = \left\|\mathcal{B}[\hat{q}(\cdot;\boldsymbol{\theta})]\right\|^2_{\partial\Omega\times(0,T]}\end{aligned} \qquad (6)$$

The relative weights, $\lambda_{PDE}$, $\lambda_{BC}$, and $\lambda_{IC}$ in Equation (6), control the trade-off between different components in the loss function. The PDE loss is computed over a finite set of $m$ collocation points $D = \{x_i, t_i\}_{i=1}^{m}$ during training, along with boundary condition loss and initial condition loss. The gradients in the loss function are computed via automatic differentiation [25].

In our previous work, to capture the local sharp transitions in the flow around airfoils, we presented NNfoil, a PINN method combined with mesh transformation. In NNfoil, neural network is used to learn the flow in computational space instead of physical space, taking $\xi, \eta$ as the input instead of $x, y$. To calculate the equation loss, we first use automatic differentiation to obtain the derivative in the computational space, then calculate the derivative in the physical space through Equation (7), and finally calculate the total loss function through Equation (6). Please refer to [24] for more details.

$$\begin{aligned} \frac{\partial}{\partial x} &= \frac{\partial}{\partial \xi}\frac{\partial \xi}{\partial x} + \frac{\partial}{\partial \eta}\frac{\partial \eta}{\partial x} \\ \frac{\partial}{\partial y} &= \frac{\partial}{\partial \xi}\frac{\partial \xi}{\partial y} + \frac{\partial}{\partial \eta}\frac{\partial \eta}{\partial y} \end{aligned} \tag{7}$$

2.3 A complete state-space solution model for the inviscid flow around airfoils

To simultaneously obtain the steady solution of the airfoil flow in the complete state-space, we expand the input dimension of NNfoil to include all state parameters, including the coordinate in computational space, Mach number, angle of attack and shape parameters. The upper and lower surfaces of the shape are parameterized by 6 Class-Shape Transform (CST) [26] parameters respectively, as elaborated in the Appendix.

We refer to the established complete state-space model as NNfoil-C. As shown in the Figure 2, the input of the model includes all state parameters. The output of the model is the velocity components $u, v$, the density $\rho$, and the pressure coefficient $C_p = (p - p_\infty)/(0.5\rho_\infty V_\infty^2)$. The model outputs pressure coefficient instead of pressure because the value of pressure changes significantly with varying Mach number, which hampers the network's optimization. The pressure coefficient serves as a physics-based normalization of pressure. Since NNfoil is based on mesh transformation, to calculate the loss function, the unit normal vector of the wall and the terms $\partial x/\partial \xi, \partial y/\partial \xi, \partial x/\partial \eta$ and $\partial y/\partial \eta$ need to be evaluated using the CST parameters in the input. To obtain these terms efficiently, the structured mesh of NACA0012 ($N_\xi \times N_\eta = 800 \times 400$) is used as the base mesh, and the meshes of all inputs are generated by mesh deformation based on the Radial Basis Function (RBF) interpolation. Since each input actually corresponds to one shape and only one grid point in computational space, we only calculate the deformed coordinates $x, y$ of the grid point and its neighbor girds, and then calculate those terms by finite difference.

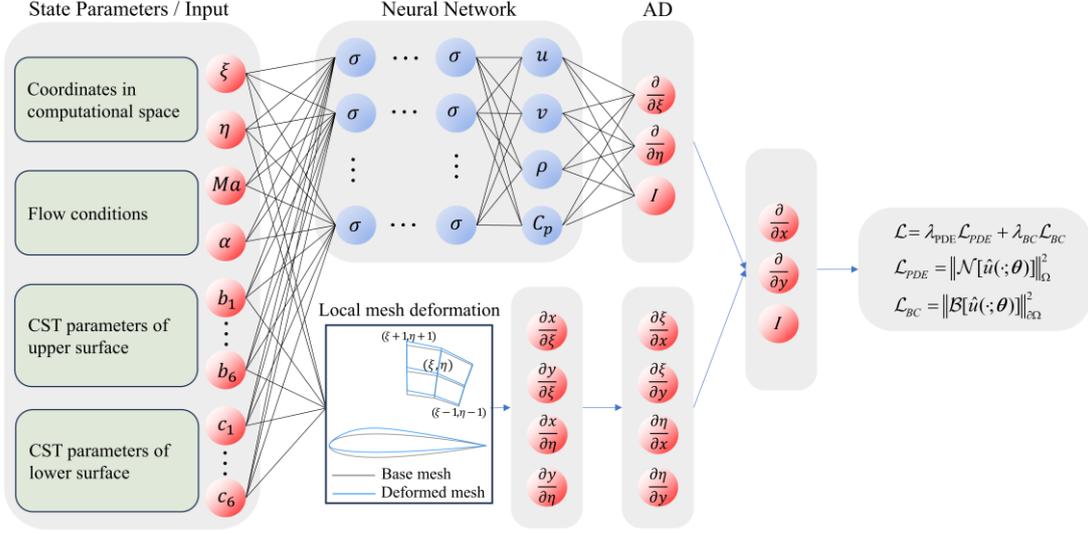

Figure 2. NNfoil-C, a complete state-space solution model for the inviscid flow around airfoils based on NNfoil.

**Pre-training**: We consider an expansive joint parameter space, denoted as $\mathcal{P}_0$ encompassing nearly all subsonic flow conditions and shapes encountered in engineering, as detailed in the next section. In the joint parameter space, we randomly sample 20,000 residual points and 2000 boundary points, and then combine them with random flow conditions and shapes as inputs. Subsequently, we employ the limited-memory Broyden-Fletcher-Goldfarb-Shanno (LBFGS) algorithm to perform 1000 iterations of gradient descent. Following this, the optimizer is restarted, new inputs are sampled, and the optimization process is repeated. This iterative procedure continues until convergence is achieved. Once the model is established, it exhibits the capability to predict the flow for diverse flow conditions and airfoils.

**Fine-tuning**: While the model can effectively derive solutions for the joint parameter space, its accuracy remains slightly inferior to NNfoil, which solves each flow individually. This discrepancy is attributed to the representation capabilities and optimization difficulties of the neural network. In many practical applications, our focus is on a specific subset $\mathcal{P}$ of the space $\mathcal{P}_0$. Inspired by the pretraining-finetuning method used in large language models, we can reset the sampling space to $\mathcal{P}$ and fine-tune the pre-trained model to achieve better performance on specific tasks. For example, to obtain a flow with specified flow condition and shape, we set the flow condition and CST parameters in the input of the pretrained model to the desired values and then continue training the model for several iterations.

## 3 Results

In this section, we present and discuss the efficiency and accuracy comparison between pre-trained and fine-tuned NNfoil-C and the classical second-order finite volume method (FVM). The FVM is one of the most widely used numerical methods in computational fluid dynamics (CFD), and it has extensive theoretical foundation for convergence. Thus, we use FVM on a very fine mesh ($N_\xi \times N_\eta = 800 \times 400$) to compute a very accurate reference solution as the ground truth. In all results, the FVM reduces the residuals of continuous equations by 10 orders of magnitude. All our FVM calculations are performed on 4 Inter i9-13900KS CPU cores, while NNfoil training are performed on an NVIDIA GeForce RTX 4090 GPU.

Throughout all benchmarks, NNfoil-C will employ the fully connected DNN architecture with 10 hidden layers and 128 neurons per hidden layer, equipped with hyperbolic tangent activation functions (tanh) and be trained using the LBFGS optimizer. We choice $\lambda_{PDE} = 2 \times 10^4, \lambda_{BC} = 1$ just like NNfoil.

### 3.1 Pre-trained model

In this subsection, we first specific the joint adoption space $\mathcal{P}_0$ and build the pre-trained model, and then compare the results of the pre-trained model and FVM on a wide range of flows. Because NNfoil is based on mesh transformation, the sampling space of coordinates $\xi, \eta$ is the grid points of the computational space. The sampling space of Mach number is $Ma \in [0.2, 0.6]$, which includes the range of subsonic flow. The sampling space of angle of attack is $\alpha \in [-5°, 5°]$, which contains the typical angle of attack of the flow around airfoils in engineering. The sampling space of CST parameters is not a continuous interval. This is because a large interval may result in non-physical airfoils, while a small interval may lead to minor shape changes. Instead, we consider the UIUC database, a rich and professional database containing approximately 1600 airfoils. We calculate the CST parameters for these airfoils. Then the sampling space of CST parameters is set to the union of 30% random perturbation space for the CST parameters of each airfoil. These practices ensure that the sampling space of CST is large enough to include almost all airfoils that may be encountered. The sampling spaces of all state parameters together form the joint parameter space $\mathcal{P}_0$, and the input of the neural network is randomly sampled in this space. We perform gradient descent using the LBFGS optimizer for $1.2 \times 10^6$ total iterations. The training takes a total of 18.8 hours, which is only 225 times that of NNfoil solving a flow (about 5min). As shown in the Figure 3, the training achieved stable convergence.

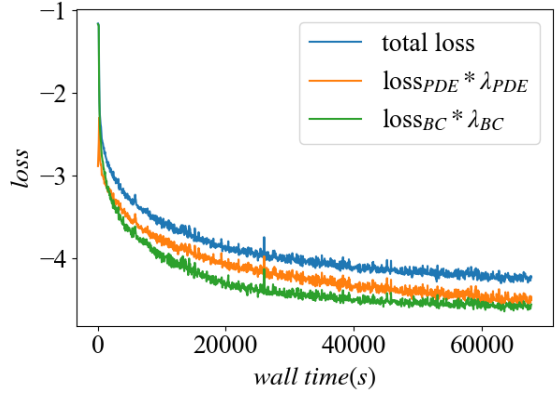

Figure 3. Training history of NNfoil-C.

We consider several classic airfoils, as shown in the Figure 4, to compare the errors of FVM and NNfoil-C. The FVM is performed on the mesh $N_\xi \times N_\eta = 200 \times 100$. We evaluate the accuracy of different methods by relative $L_1$ errors of pressure coefficient distribution $C_p = (p - p_\infty)/(0.5\rho_\infty V_\infty^2)$ on the wall, which is the most concerned in the inviscid flow around airfoils. The relative $L_1$ error between the predicted value $\hat{q}$ and the reference value $q_{ref}$ is defined as $\|\hat{q} - q_{ref}\|_1 / \|q_{ref}\|_1$.

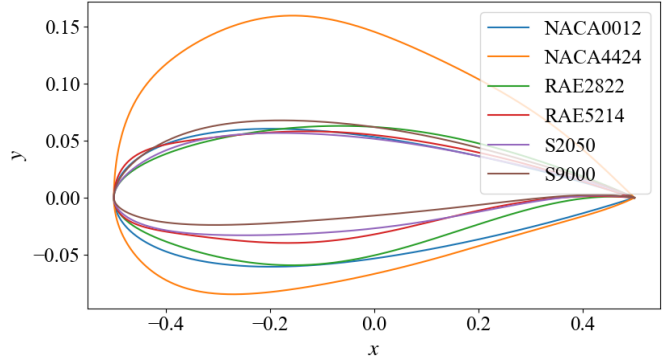

Figure 4. Airfoils used to compare the results of FVM and NNfoil-C.

The setups and results of different cases are presented in Table 1 and Figure 5. We observe that the results of the pre-trained model remain generally consistent with the reference solution over a wide range of flow conditions and airfoils. However, there are significant differences in errors between different cases, and the error in Case 5 is almost unacceptable, compromising the credibility of the model. This discrepancy can be attributed to the limited representation capacity of the fully connected neural network. Utilizing a neural network with enhanced representation capabilities and extending the training duration may potentially improve the accuracy of the results. However, the representational capacity of neural networks is always limited, and directly predicted results often lack credibility. For NNfoil-C, another alternative approach to enhancing accuracy involves rapidly fine-tuning the pre-trained model for specific tasks. This is a distinct advantage of PINN-like methods over data-driven modeling methods.

Table 1. The setups and the relative $L_1$ errors of pressure coefficient distribution for different cases.

|  | Ma | $\alpha$ | Airfoil | FVM error | Pre-trained error |
|---|---|---|---|---|---|
| Case1 | 0.2 | -5 | NACA0012 | 0.034 | 0.031 |
| Case2 | 0.3 | -3 | NACA4412 | 0.031 | 0.037 |
| Case3 | 0.4 | -1 | RAE2822 | 0.036 | 0.018 |
| Case4 | 0.5 | 1 | RAE5214 | 0.037 | 0.034 |
| Case5 | 0.6 | 3 | S2050 | 0.023 | 0.143 |
| Case6 | 0.4 | 5 | S9000 | 0.024 | 0.060 |

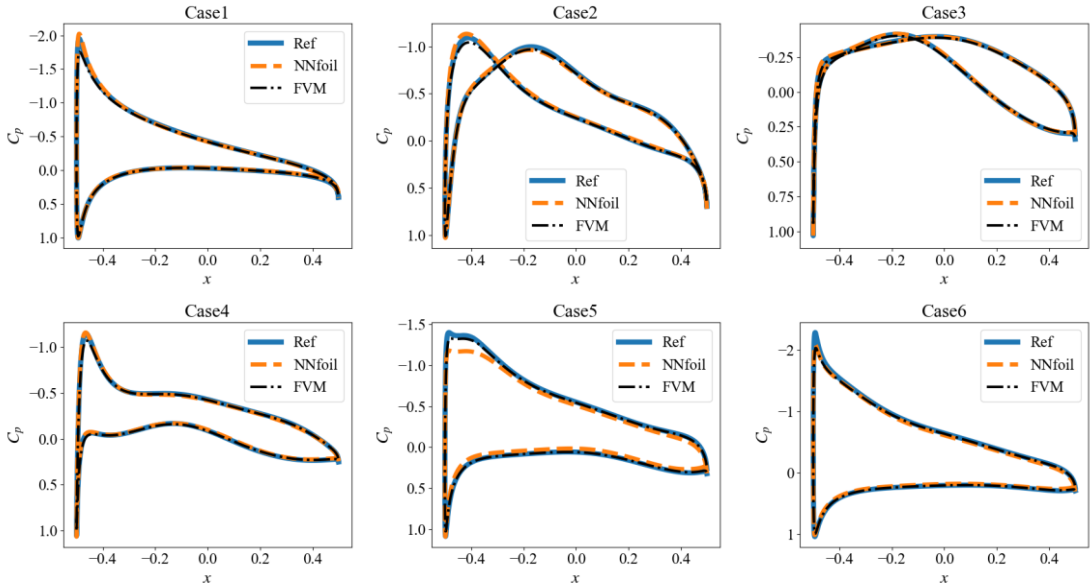

Figure 5. The pressure coefficient distributions obtained by FVM and pre-trained NNfoil-C for different cases.

### 3.2 Fine-tuned model

In this section, we fine-tune the pre-trained model for each case in the previous section to achieve higher accuracy. A popular fine-tuning method is to fine-tune a few inherent parameters while leaving the majority of parameters unchanged in model adaption. This approach does not seek to change the internal structure of a model but to optimize a small number of internal parameters to solve particular tasks, thereby obtaining good results with a small computational cost and avoiding overfitting.

We freeze the parameters of the first $n$ hidden layers of the pre-trained model separately ($n = 0, 2, \cdots, 8$) and perform 500 iterations of gradient descent using the LBFGS optimizer, where $n = 0$ indicates all parameters are trainable. Figure 6 illustrates the variation of errors over training time for the fine-tuned models with different frozen layers of the neural network. The error is the mean of 30 results, derived from the six cases in Section 3.1, with each case undergoing five random training. Unlike fine-tuning in supervised learning, fine-tuning applied to NNfoil-C does not encounter overfitting because NNfoil-C is a solution model. Therefore, maintaining all parameters trainable yields the highest accuracy, as illustrated in Figure 6.

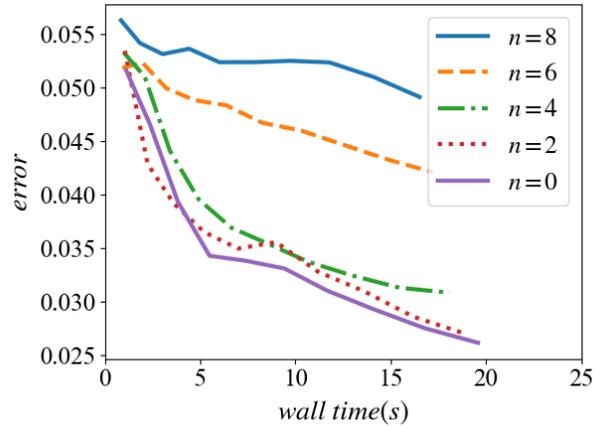

Figure 6. Average errors against wall time for different frozen layers.

As shown in Figure 7, after fine-tuning the pre-trained model, the accuracy of all results has been significantly improved, achieving an accuracy comparable to FVM. Table 2 lists detailed information about the time and accuracy of FVM and fine-tuned model. In addition, the time and accuracy of solving each flow separately using NNfoil are also listed. The fine-tuned model takes only about one-twentieth of NNfoil and one-third of FVM, but achieves comparable accuracy to them. Furthermore, NNfoil-C will demonstrate more significant advantages in a series of similar repetitive tasks, as will be shown in the next section.

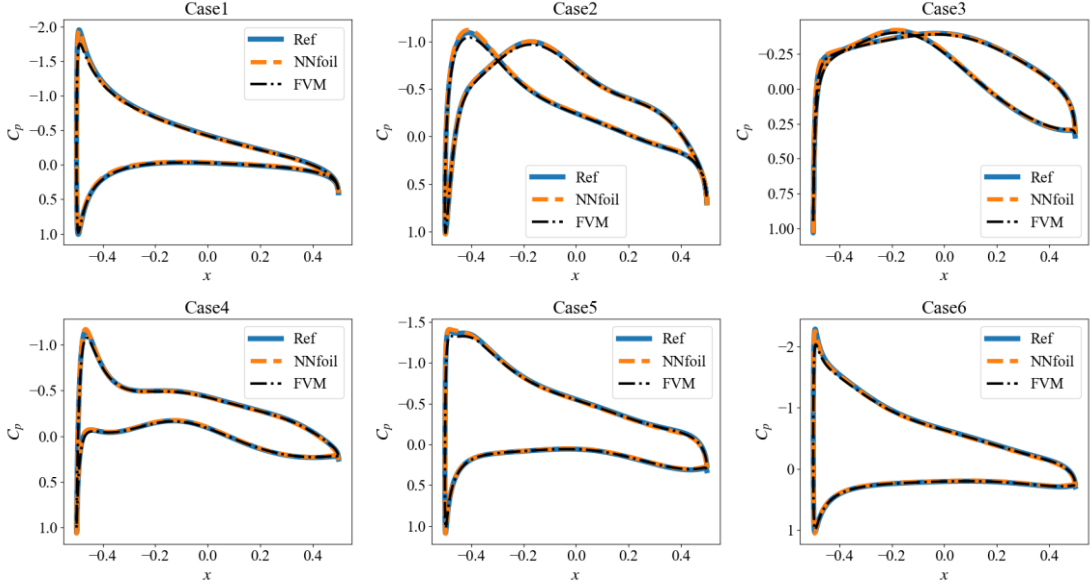

Figure 7. The pressure coefficient distributions obtained by FVM and fine-tuned NNfoil-C for different cases.

Table 2. Errors and wall time for results obtained by FVM, fine-tuned NNfoil-C, and NNfoil. This table is not meant as a rigorous comparison of the computational efficiency of the two frameworks due to many different factors involved.

|  | FVM | | Fine-tuned NNfoil-C | | NNfoil | |
|---|---|---|---|---|---|---|
|  | $C_p$ error | wall time | $C_p$ error | wall time | $C_p$ error | wall time |
| Case1 | 0.034 | 56.0s | 0.036 | 20.3s | 0.038 | 434s |
| Case2 | 0.031 | 57.3s | 0.020 | 19.9s | 0.037 | 612s |
| Case3 | 0.036 | 57.7s | 0.019 | 15.0s | 0.015 | 338s |
| Case4 | 0.037 | 74.4s | 0.027 | 20.0s | 0.022 | 265s |
| Case5 | 0.023 | 90.0s | 0.023 | 18.7s | 0.018 | 366s |
| Case6 | 0.024 | 53.6s | 0.025 | 19.9s | 0.015 | 488s |

## 4 Applications

### 4.1 Airfoil flow simulation

In aerodynamics, when assessing the performance of an airfoil, it is imperative to analyze its aerodynamic characteristics and forces under varying Mach numbers and angles of attack. In this application, instead of fine-tuning the model for each specific flow, we fine-tune the model in a new sampling space where the shape remains constant but the Mach number and the angle of attack varies. We consider the NACA2412 airfoil with $Ma \in [0.4, 0.6]$ and $\alpha \in [-5°, 5°]$.

We fine-tune the pre-trained model by performing 5000 iterations of gradient descent, taking a total of 122 seconds. The time is equivalent to the computational cost of solving the flow using FVM only twice. Figure 8 shows the results of the pre-trained model and the fine-tuned model under different Mach numbers and angles of attack. We observe that the results of the model are consistent with the reference solution in the first four examples. However, both the pre-trained and the fine-tuned models exhibit significant errors at $Ma = 0.6, \alpha = 5°$. This discrepancy is attributed to the presence of a shock wave in the flow, which the current version of NNfoil is still unable to capture. This difficulty may be solved in the future by using recent advances in the field such as [27,28]. Figure 9 shows the lift line at $Ma = 0.4$, and we observe a significant improvement through fine-tuning the pre-trained model.

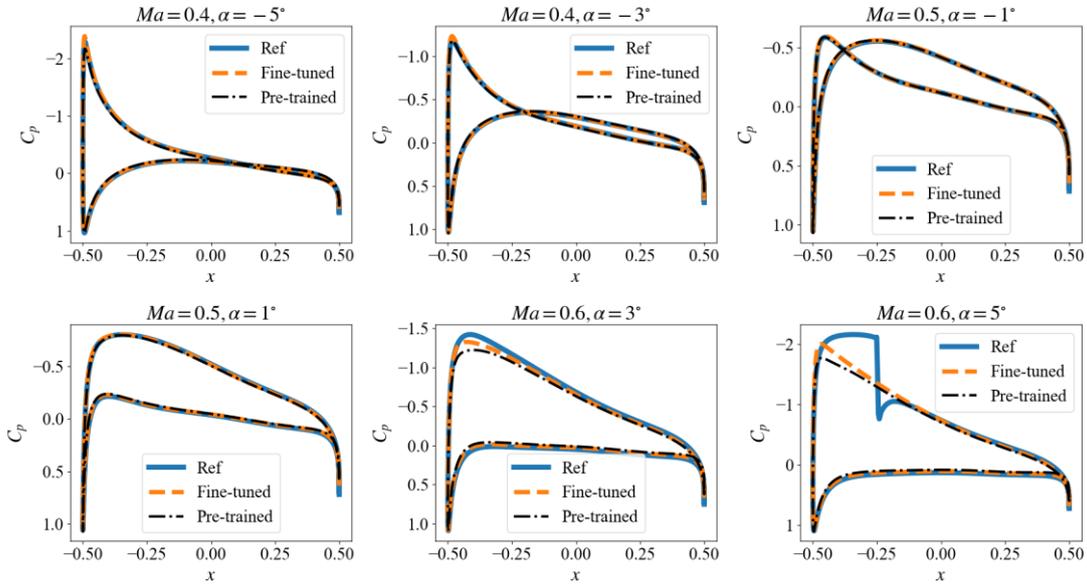

Figure 8. The pressure coefficient distributions for the NACA2412 airfoil at various Mach numbers and angles of attack.

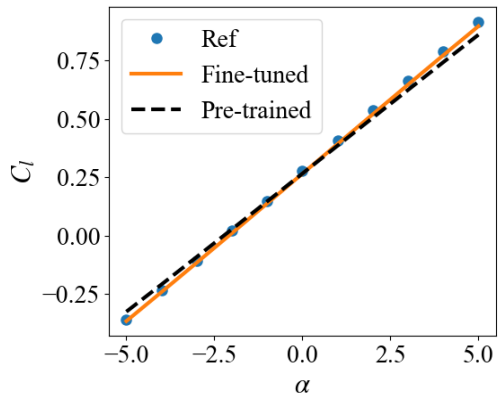

Figure 9. The lift line of the NACA2414 airfoil at Ma = 0.4.

4.2 Airfoil inverse design

Aerodynamic shape design has persistently posed a challenging issue in aerospace engineering. It endeavors to optimize airfoil shape to attain optimal performance in key physical parameters, such as maximizing lift-to-drag ratio or minimizing drag. Inverse design is a branch of aerodynamic shape design whose goal is to determine the target shape for a set of desired flow parameters. In traditional numerical methods, such optimization problems are invariably solved using some variant of gradient descent algorithm, often involving hundreds of repeated 'full solves' of the forward problem. Therefore, such tasks are often prohibitively expensive and require complex formulas and code.

Since the complete state-space solution model has been established and the model is end-to-end differentiable, we can quickly obtain the gradient of the optimization objective to the design variables (the CST parameters), thereby achieving efficient shape design optimization. We consider an inverse design task with the flow conditions $Ma = 0.4$ and $\alpha = 3°$. We aim to optimize the initial NACA0012 airfoil to align its wall pressure distribution with that of the RAE2822 airfoil, quantifying the objective through mean squared error in surface pressure distributions. Thus, the optimal shape for this task is the RAE2822 airfoil, and the optimization error is defined as the relative $L_1$ error between the current airfoil shape and the RAE2822 airfoil.

Considering the rapid evaluation capability of the pre-trained model, the first stage of optimization is to use the BFGS algorithm to obtain the optimal shape based on the pre-trained model. For any airfoil, the flow variables output from the pre-trained model are used to calculate the optimization objective, and then automatic differentiation is used to calculate the gradients of the optimization objective with respect to the design variables. In the second stage of optimization, we initialize with the design airfoil from the first stage. For any airfoil, to obtain more accurate flow and gradients of the design variables, we fine-tune the model. Instead of fine-tuning the pre-training model for each specific shape, we are constantly fine-tuning the model with the change of shape.

The history of optimization objective and optimization error as shown in Figure 10. We observe that, following the shape optimization in the first stage, the second stage converges within 5 iterations, achieving lower objective and error. As shown in Table 3, the wall time for the first stage is only 0.045 seconds, thanks to the fast predictive and end-to-end differentiable capabilities of the pre-trained model. Through fine-tuning the model, the optimization error is reduced by about 42%. Despite the significantly longer wall time of the second stage compared to the first stage, the overall cost of the inverse design is about 2.2 min, which remains much lower than traditional methods.

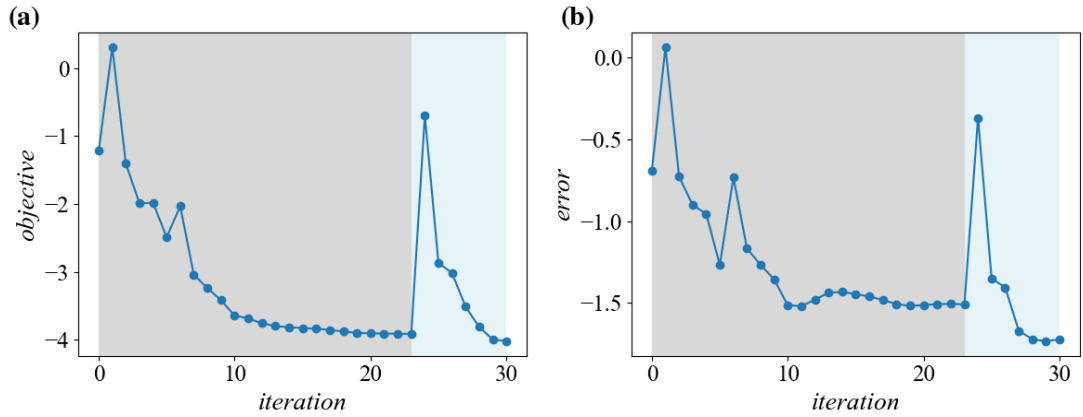

Figure 10. (a) Optimization objective and (b) optimization error with respect to iterative steps of design optimization. The grey region is the first stage, and the light region is the second stage.

Table 3. The optimization objective, optimization error, and wall time of the first stage and the second stage.

|              | objective | error | wall time |
|---|---|---|---|
| First stage  | 1.2e-4    | 0.031 | 0.05s     |
| Second stage | 9.5e-5    | 0.018 | 152s      |

Figure 11 shows the designed airfoil along with its pressure distribution. We observed that the pressure distribution of the designed airfoil is closely aligns with the target pressure distribution, but there are still slight discrepancies between the designed airfoil and the target airfoil. This is because the pressure distribution of the RAE2822 airfoil is obtained by FVM, which is not completely consistent with the pressure distribution of the fine-tuned model. Nevertheless, the precision of the designed shape remains acceptable.

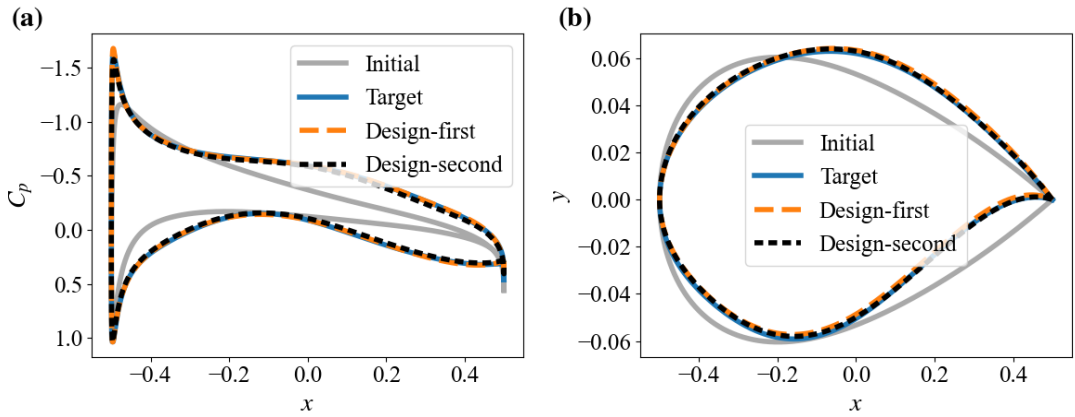

Figure 11. (a) Pressure distribution and (b) Airfoil before and after optimization.

## 5 Conclusions

In this article, we employ the parametric PINN method to construct a complete state-space solution model designed to yield solutions for subsonic inviscid flow around airfoils across a broad spectrum of flow conditions and shapes. Unlike traditional numerical methods that must solve each flow individually, the model can simultaneously obtain continuous solutions in a broad state-space, requiring only 18.8 hours of training (comparable to about 225 times the computational cost of solving a single flow). Moreover, the model allows to be fine-tuned on specific tasks, thereby quickly achieving accuracy comparable to the finite volume method. With its fast prediction capability and end-to-end differentiability, the model can complete airfoil flow simulation and shape reverse design tasks in about 1 second. As an option, we can fine-tune the model to achieve higher accuracy on both tasks, requiring only about 2 minutes. In contrast, traditional methods often require solving the flow field about hundreds of times in these tasks.

In comparison to data-driven modeling methods, the characteristics of parametric PINN method, including the utilization of unlabeled data and the ability to fine-tune the model, make it an appealing proxy modeling approach. On the one hand, it eliminates the need for extensive and costly labeled data. On the other hand, the rapid fine-tuning of the model contributes to a swift enhancement of accuracy, increasing result reliability, and reducing the reliance on the network's representational capacity. In current research, the accuracy improvement from fine-tuning is still constrained by the accuracy of the PINN method. As the accuracy and efficiency of the PINN method increase, its advantages will become more pronounced.

Despite this, we argue that parametric PINN method should not be regarded as a complete replacement for data-driven methods or traditional numerical methods. On the one hand, this method relies on accurate governing equations, which may not exist in many complex systems. On the other hand, the current generation of PINN still has limited accuracy and faces challenges in solving many complex problems. However, as knowledge about training PINN improves and hardware dedicated to neural networks becomes faster, parametric PINN methods will demonstrate superiority in addressing a broader range of problems.

## Data Availability Statement

The data that support the findings of this study are available from the corresponding author upon reasonable request.

## Conflict of Interest Statement

The authors have no conflicts to disclose.

## Acknowledgments

We would like to acknowledge the support of the National Natural Science Foundation of China (No. 92152301).

## Appendix

The Class-Shape function Transformation (CST) method is widely used in airfoil representation less design parameters and higher accuracy in geometric fitting.

The upper and lower surfaces of the airfoil can be expressed as

$$\bar{y}_u = C(\bar{x}) \cdot S(\bar{x}) + \bar{x} \cdot \Delta y_{te,u}$$
$$\bar{y}_l = C(\bar{x}) \cdot S(\bar{x}) + \bar{x} \cdot \Delta y_{te,l} \quad \text{(A-1)}$$

where $\Delta y_{te,u}$ and $\Delta y_{te,l}$ are the airfoil trailing edge thickness, they are zero in this paper. The class function and shape function are defined as

$$C(\bar{x}) = (\bar{x})^{N_1}(1-\bar{x})^{N_2}, 0 \leq \bar{x} \leq 1$$
$$S(\bar{x}) = \sum_{i=0}^{n} A_i \cdot S_i(\bar{x}) = \sum_{i=0}^{n} A_i \cdot \frac{n!}{i!(n-i)!} \cdot (\bar{x})^i (1-\bar{x})^{n-i} \quad \text{(A-2)}$$

where different combinations of the exponents $N_1$ and $N_2$ in the class function define a variety of basic general classes of geometric shapes. For a round-nose airfoil $N_1 = 0.5$ and $N_2 = 1.0$. $n$ is the order of Bernstein polynomial. $A_i$ is the design parameter.